\documentclass[useAMS,usenatbib]{mn2e}
\usepackage{epsfig}
\usepackage{deluxetable} 

\title[High-energy properties of the FSRQ PKS 2149$-$306]{High-energy properties of the high-redshift flat spectrum radio quasar PKS 2149$-$306}
\author[F. D'Ammando, M. Orienti]{F. D'Ammando$^{1,2}$\thanks{E-mail: dammando@ira.inaf.it}, M. Orienti$^{2}$\\
$^{1}$Dip. di Fisica e Astronomia , Universit\`a di Bologna, Viale Berti Pichat 6/2, I-40127 Bologna, Italy \\
$^{2}$INAF - Istituto di Radioastronomia, Via Gobetti 101, I-40129 Bologna, Italy\\}
\begin{document}

\date{Accepted. Received; in original form}

\maketitle

\label{firstpage}

\begin{abstract}
We investigate the $\gamma$-ray and X-ray properties of the flat spectrum
radio quasar PKS 2149$-$306 at redshift $z$ = 2.345. A strong $\gamma$-ray flare from this source was detected by the Large Area
Telescope on board the {\em Fermi Gamma-ray Space Telescope} satellite in 2013 January, reaching on January 20 a daily
peak flux of (301 $\pm$ 36)$\times$10$^{-8}$ ph cm$^{-2}$ s$^{-1}$ in the
0.1--100 GeV energy range. This flux corresponds to an apparent isotropic luminosity of (1.5$\pm$0.2)$\times$10$^{50}$ erg
s$^{-1}$, comparable to the highest values observed by a blazar so far. During
the flare the increase of flux was accompanied by a significant change of the
spectral properties. Moreover significant flux variations on a 6-h time-scale
were observed, compatible with the light crossing time of the event horizon of
the central black hole.

\noindent The broad band X-ray spectra of PKS 2149$-$306 observed by {\em Swift}-XRT and {\em NuSTAR} are well described by a broken power-law model,
with a very hard spectrum ($\Gamma_{1}$ $\sim 1$) below the break energy, at
E$_{\rm\,break}$ = 2.5--3.0 keV, and $\Gamma_{2}$ $\sim 1.4$--1.5 above the
break energy. The steepening of the spectrum below $\sim 3$ keV may indicate
that the soft X-ray emission is produced by the low-energy relativistic electrons. This is in agreement with the small variability amplitude and the
lack of spectral changes in that part of the X-ray spectrum observed between
the two {\em NuSTAR} and {\em Swift} joint observations. As for the other
high-redshift FSRQ detected by both {\em Fermi}-LAT and {\em Swift}-BAT, the photon
index of PKS 2149$-$306 in hard X-ray is 1.6 or lower and the average
$\gamma$-ray luminosity higher than 2$\times$10$^{48}$ erg s$^{-1}$.   
\end{abstract} 


\section{Introduction}

Blazars are radio-loud active galactic nuclei (AGN), with powerful relativistic
jets observed at a small viewing angle. For this reason their emission is strongly
enhanced due to Doppler boosting and they are expected to be detected up to high
redshift. The most distant blazar identified so far is Q0906$+$6930
\citep{romani04}, located at redshift $z$ = 5.47. A possible excess of
$\gamma$-ray photons from a position compatible with this flat spectrum radio
quasar (FSRQ) was observed by EGRET \citep{romani06}, but has not been
confirmed by {\em Fermi} Large Area Telescope (LAT) observations so far. Recently, two FSRQ at redshift $z$ $>$
5, B2 1023$+$25 and SDSS J114657.79$+$403708.6, were detected in hard
X-rays by {\em NuSTAR} and identified as blazars \citep{sbarrato13, ghisellini14a}. Both
objects have never been detected in $\gamma$ rays. The most distant FSRQ
reported in the Third {\em Fermi}-LAT catalogue \citep[3FGL;][]{acero15} is PKS
0537$-$286 at redshift $z$ = 3.104, indicating the difficulty in detecting quasars at $z$ $> 3$ in the $\gamma$-ray regime.
  
PKS 2149$-$306 \citep[RA = 21h51m55.5239s, Dec. = $-$30$^{\circ}$27\arcmin53\farcs697, J2000;][]{johnston95} is a FSRQ at redshift $z = 2.345$ \citep{wilkes86}. The source is bright in X-rays, showing substantial variability both in intensity and
spectral slope, as indicated by {\em ROSAT} \citep{siebert96}, {\em ASCA} \citep{cappi97}, {\em XMM-Newton}
\citep{ferrero03}, and {\em Swift} observations \citep{sambruna07,bianchin09}. A tentative detection of an emission line at
$\sim$17 keV in the source frame by {\em ASCA} was interpreted as highly-blueshifted Fe K$\alpha$ \citep{yaqoob99}. This finding was not
confirmed by \citet{fang01} and \citet{page04} using {\em Chandra} data. As
for other FSRQ, a low-energy photon deficit in X-rays was suggested for PKS
2149$-$306, possibly due to an absorbing cloud in the source rest frame
\citep[e.g.,][]{sambruna07} or to a low-energy tail of the electron population
\citep[e.g.,][]{tavecchio07}. PKS 2149$-$306 was detected in hard X-rays with a hard spectrum by {\em Beppo}SAX \citep{elvis00}, {\em Swift}-BAT
\citep{baumgartner13}, {\em INTEGRAL}-IBIS \citep{beckmann09}, and lately {\em NuSTAR} \citep{tagliaferri15}. 

Among the high-redshift ($z > 2$) blazars, 64 were reported in the Third {\em Fermi} LAT Catalog \citep[3FGL;][]{acero15}. Only two of these objects are at redshift $z > 3$. In contrast, ten blazars at redshift $z > 3$ were detected in hard X-rays by {\em Swift}-BAT \citep{baumgartner13}, {\em INTEGRAL}-IBIS \citep{bassani12} and {\em NuSTAR} \citep{sbarrato13,ghisellini14a}. In particular, seven blazars at redshift $z > 3$ are detected by {\em Swift}-BAT. Therefore observations in the hard X-ray band seem to be more effective than the $\gamma$-ray band for finding blazars at redshift $z > 3$. This might be due to the fact that high-redshift blazars generally have the inverse Compton (IC) peak at hundreds of keV and thus are not ideal for a detection at GeV energies \citep[e.g.,][]{ghisellini11,ghisellini13}. For this reason, the detection of a $\gamma$-ray flare from a high-redshift blazar may be even more interesting with respect to the flaring activity from other blazars. 

\noindent On 2013 January 4 a strong $\gamma$-ray flare from PKS 2149$-$306 was detected by {\em Fermi}-LAT \citep[preliminary results were
reported in][]{dammando13}. The aim of this paper is to discuss the $\gamma$-ray and X-ray properties of this source and to make a comparison with
the other high-redshift blazars detected by {\em Fermi}-LAT and {\em Swift}-BAT using the 3FGL catalogue \citep{acero15} and the 70-month {\em Swift}-BAT catalogue \citep{baumgartner13}.

This paper is organized as follows. In Section 2 we report the LAT data analysis and
results, while in Sections 3, 4, and 5 we present the results of the {\em Swift}, {\em XMM-Newton}, and {\em NuSTAR} observations, respectively. We discuss the properties of the source in Section 6, while in
Section 7 we summarize our results. Throughout the paper, a
$\Lambda$--cold dark matter cosmology with $H_0$ = 71 km s$^{-1}$ Mpc$^{-1}$, $\Omega_{\Lambda} = 0.73$, and $\Omega_{m} =
0.27$ is adopted \citep{komatsu11}. The corresponding luminosity distance at $z = 2.345$
(i.e. the source redshift) is d$_L =  19240$\ Mpc. Throughout this paper the quoted uncertainties are
given at 1$\sigma$ level, unless otherwise stated. For power law spectra $dN/dE \propto E^{-\Gamma_{\rm\,X,\gamma}}$ we denote as $\Gamma_{\rm\,X}$
and $\Gamma_{\gamma}$ the spectral indices in the X-ray and $\gamma$-ray bands, respectively.

\section{{\em Fermi}-LAT Data: Selection and Analysis}
\label{FermiData}

The {\em Fermi}-LAT  is a pair-conversion telescope operating from 20 MeV to
more than 300 GeV. It has a large peak effective area ($\sim 8000$ cm$^{2}$ for 1
GeV photons), and a field of
view of about 2.4 sr with an angular resolution (68 per cent
containment angle) of 0\fdg6 for a single-photon at $E$ = 1 GeV on-axis. Details about the {\em Fermi}-LAT are given in \citet{atwood09}. 

The LAT data reported in this paper were collected from 2008 August 4  (MJD
54682) to 2014 August 4 (MJD 56873). During this period the LAT instrument
operated almost entirely in survey mode. The analysis was performed with the
\texttt{ScienceTools} software package version v9r33p0\footnote{http://fermi.gsfc.nasa.gov/ssc/data/analysis/software/}. Only events belonging
to the `Source' class were used. In addition, a cut on the zenith angle ($< 100^{\circ}$) was applied to reduce contamination from
the Earth limb $\gamma$ rays, which are produced by cosmic rays interacting with the upper atmosphere. 
The spectral analysis was performed with the instrument response functions \texttt{P7REP\_SOURCE\_V15} using an unbinned maximum likelihood method implemented in the Science tool \texttt{gtlike}. Isotropic (`iso\_source\_v05.txt') and Galactic diffuse
emission (`gll\_iem\_v05\_rev1.fit') components were used to model the
background\footnote{http://fermi.gsfc.nasa.gov/ssc/data/access/lat/\\BackgroundModels.html}. The
normalisations of both components were allowed to vary freely during the spectral fittings.

We analysed a region of interest of $10^{\circ}$ radius centred at the
location of PKS 2149$-$306. We evaluated the significance of the $\gamma$-ray signal from the source by
means of a maximum-likelihood test statistic TS = 2$\times$(log$L_1$ $-$ log$L_0$), where
$L$ is the likelihood of the data given the model with ($L_1$) or without
($L_0$) a point source at the position of PKS 2149$-$306
\citep[e.g.,][]{mattox96}. The source model used in
\texttt{gtlike} includes all the point sources from the 3FGL catalogue that
fall within $15^{\circ}$ of PKS 2149$-$306. The spectra of these sources
were parametrized by a power law (PL), a log parabola (LP), or a super
exponential cut-off, as in the 3FGL catalogue. 
A first maximum likelihood analysis was performed to remove from the model the sources having
TS $< 10$ and/or the predicted number of counts based on the fitted model
$N_{\rm\,pred} < 1 $. A second maximum likelihood analysis was performed
on the updated source model. In the fitting procedure, the normalization
factors and the spectral shape parameters of the
sources lying within 10$^{\circ}$ of PKS 2149$-$306 were left as free
parameters. For the sources located between 10$^{\circ}$ and 15$^{\circ}$ from
our target, we kept the normalization and the spectral shape parameters fixed to the values
from the 3FGL catalogue. 

Integrating over the period 2008 August 4--2014 August 4 (MJD 54682--56873) using a PL model, $dN/dE \propto$
$(E/E_{0})^{-\Gamma}$, the fit results in TS = 2096 in the 0.1--100 GeV energy
range, and a photon index $\Gamma_{\gamma}$ = 2.79 $\pm$ 0.03. The average flux is (10.6 $\pm$ 0.4)$\times$10$^{-8}$ ph
cm$^{-2}$ s$^{-1}$. In order
to test for curvature in the $\gamma$-ray spectrum of PKS 2149$-$306 an
alternative spectral model to a PL, a LP, $dN/dE \propto$ $(E/E_{0})^{-\alpha-\beta\,log(E/E_{0})}$, was used for the fit.
We obtain a spectral slope $\alpha$ = 2.36 $\pm$ 0.05 at the reference energy
$E_0$ = 221 MeV, a curvature parameter around the peak $\beta$ = 0.29 $\pm$ 0.03, with TS = 2183 and an average flux of
(9.7 $\pm$ 0.4)$\times$10$^{-8}$ ph cm$^{-2}$ s$^{-1}$ (Table~\ref{LAT_table}). We used a likelihood ratio test (LRT) to check the PL model (null hypothesis) against the LP
model (alternative hypothesis). Following \citet{nolan12} these values may be
compared by defining the curvature Test Statistic TS$_{\rm curve}$=TS$_{\rm
  LP}$--TS$_{\rm PL}$, which in this case results in TS$_{\rm curve}$ = 87 ($\sim 9.3\,\sigma$), meaning that we have evidence of significant curvature in the average $\gamma$-ray spectrum. 

\begin{figure}
\centering
\includegraphics[width=7.5cm]{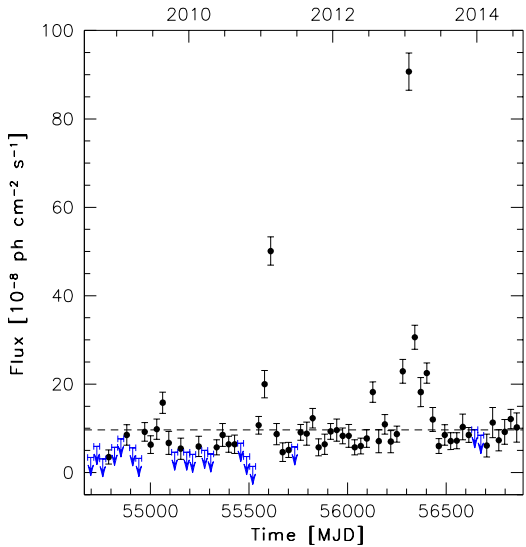}
\caption{Integrated flux light curve of PKS 2149$-$306 obtained by {\em Fermi}-LAT in the 0.1--100 GeV energy range during 2008 August 4--2014 August 4 (MJD 54682--56873) using a LP model with 30-day time bins. Arrows refer to 2$\sigma$ upper limits on the source flux. Upper limits are computed when TS $<$ 10. The dashed line represents the mean flux.}
\label{Fig1}
\end{figure}

Fig.~\ref{Fig1} shows the $\gamma$-ray light curve for the first 6 years of
{\em Fermi}-LAT observations of PKS 2149$-$306 using a LP model and 1-month
time bins. For each time bin, the spectral shape parameters of PKS 2149$-$306 and
all sources within 10$^{\circ}$ of it were frozen to the values resulting from
the likelihood analysis over the entire period. If TS $<$ 10, 2$\sigma$ upper
limits were calculated. The statistical uncertainty in the fluxes are larger
than the systematic uncertainty \citep{ackermann12} and only the former is considered in the paper. 

During the first two years of {\em Fermi} operation, PKS 2149$-$306 was
observed in a low-activity state, with an average (0.1--100 GeV) flux of (6.4
$\pm$ 0.6)$\times$10$^{-8}$ ph cm$^{-2}$ s$^{-1}$ and a spectrum described by
a PL model with photon index of $\Gamma_{\gamma}$ = 3.00 $\pm$ 0.09 \citep{nolan12}. Different from the
average spectrum over 6 years of observations, no significant curvature of the
spectrum was observed during the low-activity period between 2008 August--2010
July. This difference may be related to the low statistics in the first two
years that may prevent the detection of spectral curvature, as observed for
other FSRQ \citep[e.g., PKS 1510$-$089 and S5 0836$+$710;][]{abdo10b,
  akyuz13}. Flaring activity from this source was first observed in 2011 February, and subsequently an even stronger flare was detected in 2013 January (Fig.~\ref{Fig1}). 
  
\begin{table*}
\caption{Unbinned likelihood spectral fit results.}
\begin{tabular}{lccc|ccccc}
 \hline 
&  &  \multicolumn{2}{c}{PL} & \multicolumn{3}{c}{LP} & \\
Date (UT) & Date (MJD) & $\Gamma$ & TS$_{\rm PL}$ & $\alpha$
&$\beta$  & TS$_{\rm LP}$ & TS$_{\rm Curve}$ \\
\hline
2008-Aug-04/2014-Aug-04 & 54682--56873 & 2.79 $\pm$ 0.03 & 2096 & 2.36 $\pm$ 0.05 & 0.29 $\pm$ 0.03 & 2183 & 87\\       
2011-Feb-03/2011-Mar-05 & 55595--55625 & 2.85 $\pm$ 0.08 &  521 & 2.53 $\pm$ 0.13 & 0.28 $\pm$ 0.09 & 538  & 18\\
2013-Jan-04/2013-Feb-02 & 56296--56325 & 2.45 $\pm$ 0.05 & 1239 & 1.99 $\pm$ 0.11 & 0.28 $\pm$ 0.06 & 1273 & 34\\
\hline
  \end{tabular}
\label{LAT_table}
\end{table*} 

\subsection{Flaring periods}

Leaving the spectral shape parameters free to vary during the first high-activity period (2011 February 3--March 5; MJD 55595--55625), using a LP
model, the fit results in a spectral slope $\alpha$ = 2.53 $\pm$ 0.13 at the
reference energy $E_0$ = 221 MeV, a curvature parameter around the peak
$\beta$ = 0.28 $\pm$ 0.09, with TS = 538 and an average flux of (51.3 $\pm$
4.1)$\times$10$^{-8}$ ph cm$^{-2}$ s$^{-1}$. Using a PL model, the fit results
in TS = 521 and a photon index of $\Gamma_{\gamma}$ = 2.85 $\pm$ 0.08 (Table~\ref{LAT_table}). Using an LRT, we obtain TS$_{\rm\,curve}$ = 18 ($\sim 4.2\,\sigma$), i.e. a significant curvature of the $\gamma$-ray spectrum in 2011 February.

During the second high activity period (2013 January 4--February 2; MJD 56296--56325), using a LP model the fit
results in a spectral slope $\alpha$ = 1.99 $\pm$ 0.11 at the reference energy
$E_0$ = 221 MeV, a curvature parameter around the peak $\beta$ = 0.28 $\pm$
0.06, with TS = 1273 and an average flux of (77.4 $\pm$ 5.1)$\times$10$^{-8}$
ph cm$^{-2}$ s$^{-1}$. Using a PL model the fit results in TS = 1239 and a
photon index of $\Gamma_{\gamma}$ = 2.45 $\pm$ 0.05 (Table
\ref{LAT_table}). Using an LRT, we obtain TS$_{\rm curve}$ = 34 ($\sim 5.8\,\sigma$), indicating a significant curvature of the $\gamma$-ray spectrum in that period.

\begin{figure*}
\begin{center}
\rotatebox{0}{\resizebox{!}{87mm}{\includegraphics{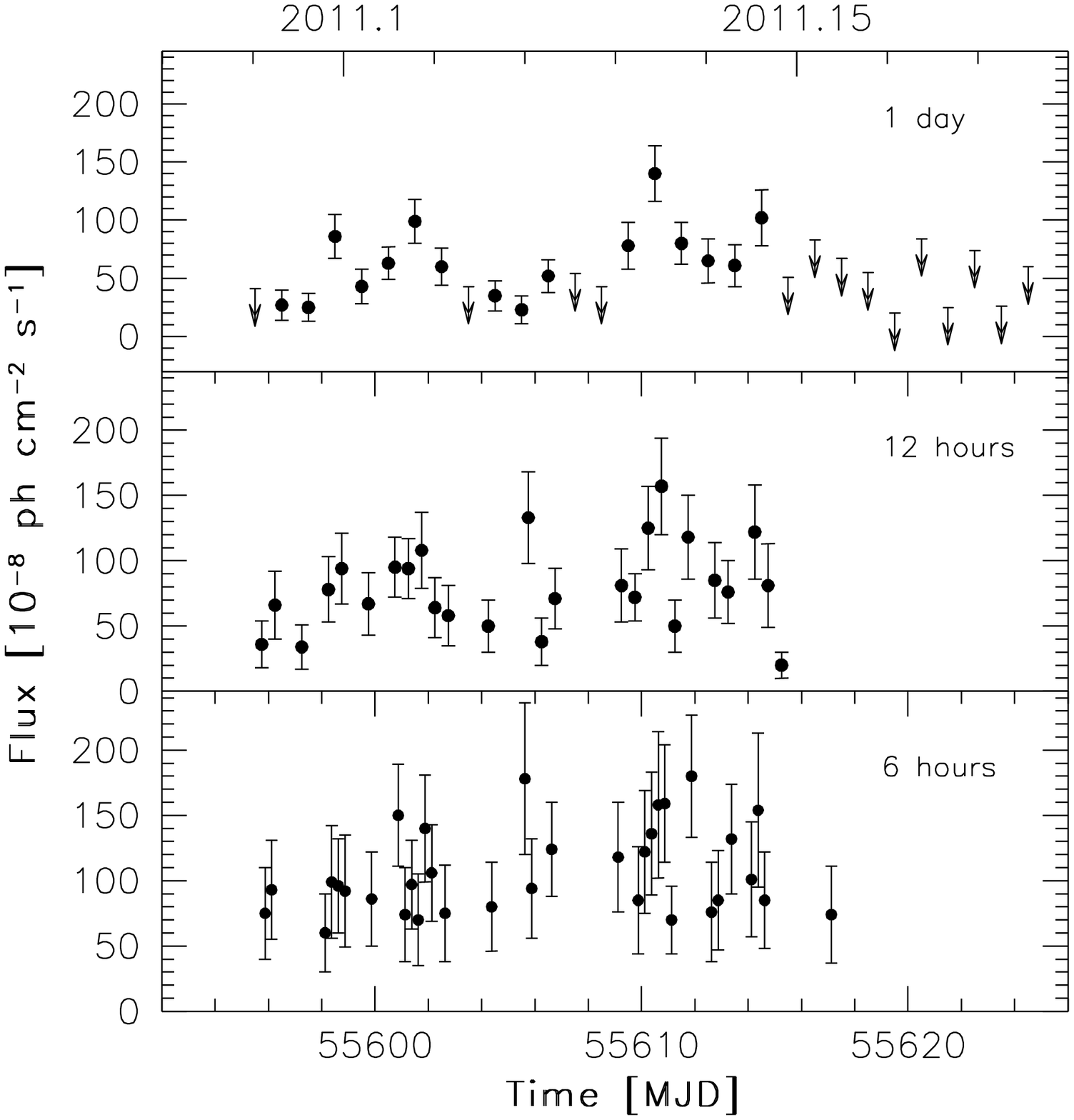}}}
\rotatebox{0}{\resizebox{!}{87mm}{\includegraphics{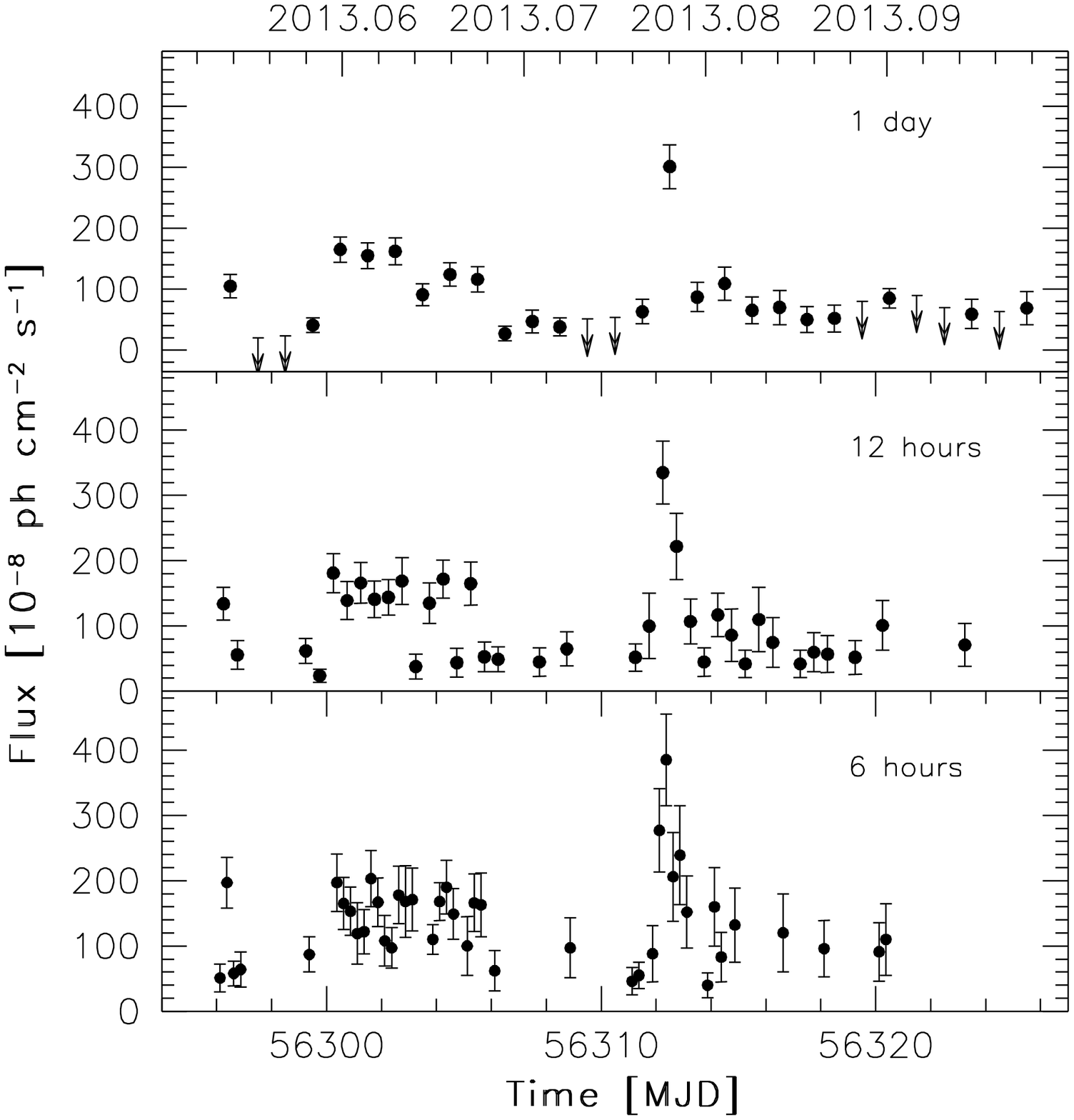}}}
\caption{\small{Integrated flux light curve of PKS 2149$-$306 obtained by {\em Fermi}-LAT in the 0.1--100 GeV energy range during 2011 February 3--March 5 ({\it left plot}) and 2013 January 4--February 2 ({\it right plot}), with 1-day time bins (upper panel), 12-h time bins (middle panel), and 6-h time bins (bottom panel). Arrows refer to 2$\sigma$ upper limits on the source flux. Upper limits are computed when TS $<$ 10. In the middle and bottom panels upper limits are not shown.}}
\label{LAT_flare}
\end{center}
\end{figure*}

In the following analysis of the light curves on sub-daily time-scales, we fixed the flux of the diffuse emission components at the value obtained by fitting the data over the respective daily time-bins. In Fig.~ \ref{LAT_flare}
we show a light curve focused on the period 2011 February 3--March 5 (left
plot) and 2013 January 4--February 2 (right plot), with 1-day (upper panel), 12-h (middle panel), and 6-h (lower panel) time bins. For each time bin, the spectral shape parameters of PKS 2149$-$306 and
all sources within 10$^{\circ}$ of it were frozen to the values resulting from
the likelihood analysis over the entire period considered.

In 2011 the daily peak of the emission was observed on February 18 (MJD 55610) with a flux of (140 $\pm$ 24)$\times$10$^{-8}$ ph cm$^{-2}$ s$^{-1}$ in the 0.1--100 GeV energy range, a factor of about 14 higher than the average flux over 6 years of {\em Fermi} observations. The
corresponding apparent isotropic $\gamma$-ray luminosity peak in the 0.1--100
GeV energy range is (5.3 $\pm$ 0.9)$\times$10$^{49}$ erg s$^{-1}$. On 12-h and 6-h time-scale the
observed peak flux is (157 $\pm$ 37)$\times$10$^{-8}$ and (180 $\pm$ 47)$\times$10$^{-8}$ ph cm$^{-2}$
s$^{-1}$, corresponding to an apparent isotropic $\gamma$-ray luminosity of
(5.9 $\pm$ 1.3)$\times$10$^{49}$ and (6.8 $\pm$ 1.7)$\times$10$^{49}$ erg s$^{-1}$, respectively.

In 2013 the daily peak of the emission was observed in January 20 (MJD 56312) with a flux of (301 $\pm$ 36)$\times$10$^{-8}$ ph cm$^{-2}$ s$^{-1}$ in the 0.1--100 GeV energy range, i.e. a factor of about 30 higher than the average flux over 6 years of {\em Fermi} observations. The corresponding apparent isotropic $\gamma$-ray luminosity peak in the 0.1--100 GeV energy range is (1.5 $\pm$ 0.2)$\times$10$^{50}$ erg s$^{-1}$. On 12-h and 6-h time-scales the
observed peak flux is (335 $\pm$ 48)$\times$10$^{-8}$ and (385 $\pm$ 70)$\times$10$^{-8}$ ph cm$^{-2}$ s$^{-1}$, corresponding to an apparent
isotropic $\gamma$-ray luminosity of (1.6 $\pm$ 0.2)$\times$10$^{50}$ and (1.9 $\pm$ 0.3)$\times$10$^{50}$erg s$^{-1}$, respectively. By means of the
\texttt{gtsrcprob} tool we estimated that during this flare the highest energy photon emitted by PKS 2149$-$306 (with probability $>$ 80 per cent to be associated with the target) was observed on 2013 January 12 with an energy of 4.8 GeV.

\begin{table*}
\caption{Log and fitting results of {\em Swift}-XRT observations of PKS
  2149$-$306 using a PL model with $N_{\rm H}$ fixed to the Galactic
  absorption. $^{(*)}$The model that best fit this observation is a broken power-law (see Section \ref{XRTdata}).}
\begin{center}
\begin{tabular}{cccccc}
\hline
\multicolumn{1}{c}{Date} &
\multicolumn{1}{c}{Date} &
\multicolumn{1}{c}{Net Exposure Time} &
\multicolumn{1}{c}{Photon index} &
\multicolumn{1}{c}{Flux 0.3--10 keV} &
\multicolumn{1}{c}{$\chi^2$/d.o.f.}\\
\multicolumn{1}{c}{(UT)} &
\multicolumn{1}{c}{(MJD)} &
\multicolumn{1}{c}{(s)} &
\multicolumn{1}{c}{($\Gamma_{\rm\,X}$)} &
\multicolumn{1}{c}{($\times$10$^{-11}$ erg cm$^{-2}$ s$^{-1}$)} &
\multicolumn{1}{c}{}\\
\hline
2005-Dec-10 & 53714 & 3314 & $1.47 \pm 0.08$ & $1.52 \pm 0.12$ & 44/45 \\
2005-Dec-13 & 53717 & 2255 & $1.39 \pm 0.10$ & $1.69 \pm 0.10$ & 32/25 \\
2009-Apr-23 & 54944 & 3114 & $1.23 \pm 0.11$ & $1.32 \pm 0.07$ & 36/31 \\
2009-Apr-29 & 54950 & 1773 & $1.34 \pm 0.14$ & $1.21 \pm 0.09$ & 16/16 \\
2009-May-05 & 54956 & 2889 & $1.36 \pm 0.10$ & $1.22 \pm 0.07$ & 28/28 \\
2009-May-14 & 54965 & 2924 & $1.32 \pm 0.08$ & $1.75 \pm 0.07$ & 40/41 \\
2009-May-23 & 54974 & 2989 & $1.40 \pm 0.11$ & $1.58 \pm 0.09$ & 31/28 \\
2009-May-29 & 54980 & 2565 & $1.19 \pm 0.08$ & $1.91 \pm 0.10$ & 39/36 \\
2010-May-11 & 55327 & 4760 & $1.30 \pm 0.08$ & $1.17 \pm 0.05$ & 38/44 \\
2011-May-07 & 55688 & 2947 & $1.01 \pm 0.09$ & $1.96 \pm 0.12$ & 35/38 \\
2011-Nov-10 & 55875 & 3261 & $1.33 \pm 0.07$ & $1.87 \pm 0.09$ & 39/48 \\
2013-Dec-16/17 & 56642/43 & 7956 & $1.09 \pm 0.04$ & $2.80 \pm 0.08$ & 143/139 \\
2014-Mar-28 & 56744 & 2537 & $1.13 \pm 0.08$ & $2.58 \pm 0.11$ & 47/44 \\
2014-Apr-18 & 56765 & 6331 & $1.08 \pm 0.05$ & $2.60 \pm 0.07$ & 126/106$^{*}$\\
\hline
\end{tabular}
\end{center}
\label{XRT}
\end{table*} 

\section{{\em Swift} Data: Analysis and Results}
\label{SwiftData}

The {\em Swift} satellite \citep{gehrels04} performed sixteen observations of PKS 2149$-$306 between 2005 December and 2014 April. The observations were performed with all three instruments on board: the X-ray Telescope \citep[XRT;][0.2--10.0 keV]{burrows05}, the Ultraviolet/Optical Telescope \citep[UVOT;][170--600 nm]{roming05} and the Burst Alert Telescope \citep[BAT;][15--150 keV]{barthelmy05}.

\subsection{{\em Swift}-BAT}

The hard X-ray flux of this source is below the sensitivity of the BAT instrument for the short exposures of the single observations, therefore those data from this instrument were not used. On the other hand, the source is included in the {\em Swift} BAT 70-month hard X-ray
catalogue \citep{baumgartner13}. The 14--195 keV spectrum is well described by a power law with photon index of $\Gamma_{\rm\,X}$ = 1.50 $\pm$ 0.10 ($\chi$$^2$/d.o.f. = 4.8/6). The resulting 14--195 keV flux is (8.3 $\pm$ 0.6)$\times$10$^{-11}$ erg cm$^{-2}$ s$^{-1}$.

\subsection{{\em Swift}-XRT}\label{XRTdata}

The XRT data were processed with standard procedures, filtering, and screening
criteria by using the \texttt{xrtpipeline v0.13.0} included in the
\texttt{HEASoft} package (v6.15)\footnote{http://heasarc.nasa.gov/lheasoft/}. The data were collected in
photon counting mode for all the observations. The source count rate
was low ($<$ 0.5 counts s$^{-1}$); thus pile-up correction was not
required. The data collected in observations separated by less than
twenty-four hours (i.e. 2010 May 11, obsid: 31404008 and 31404009; 2011 May 7, obsid: 31404010 and 31404011; 2013 December 16--17, obsid: 31404013 and 31404014) were
summed in order to have enough statistics to obtain a good spectral fit. Source events were extracted from a circular region with a radius of
20 pixels \citep[1 pixel $\sim$ 2.36 arcsec;][]{burrows05}, while background events were extracted
from a circular region with radius of 50 pixels far away from bright sources. Ancillary response files were generated with \texttt{xrtmkarf}, and
account for different extraction regions, vignetting and point spread function
corrections. We used the spectral redistribution matrices in the
Calibration database (CALDB) maintained by HEASARC\footnote{https://heasarc.gsfc.nasa.gov/}. The spectra were rebinned with a
minimum of 20 counts per energy bin to allow for $\chi^{2}$ spectrum
fitting. Bad channels, including zero-count bins, were ignored in the fit. We
have fitted the spectrum using \texttt{Xspec} \footnote{https://heasarc.gsfc.nasa.gov/xanadu/xspec/manual/manual.html}  with an absorbed power-law using the photoelectric absorption model
\texttt{tbabs} \citep{wilms00}, with a neutral hydrogen column density fixed to its Galactic value \citep[1.63$\times$10$^{20}$
cm$^{-2}$;][]{kalberla05}.
The results are reported in Table~\ref{XRT}. All errors are given at the 90\%
confidence level. Symmetric errors are reported, obtained by averaging the positive and negative errors calculated with \texttt{Xspec}.

For the observations performed on 2013 December 16--17 and 2014 April 18 there
is enough statistic for testing a more detailed spectral model with respect to
a simple power law. For the 2013 December observations, using a broken
power-law the fit results in $\Gamma_{1}$ = 0.95$^{+0.08}_{-0.13}$ below the
break energy $E_{\rm\,break}$ = 2.40$^{+0.64}_{-0.92}$ keV and $\Gamma_{2}$ =
1.32$^{+0.16}_{-0.15}$ above $E_{\rm\,break}$. The fit with a broken power-law ($\chi^{2}$/d.o.f =
129/137) does not improve with respect to a simple power law ($\chi^{2}$/d.o.f = 143/139). 
For the 2014 April observation using a broken power-law the fit results in
$\Gamma_{1}$ = 0.97 $\pm$ 0.09 below the break energy $E_{\rm\,break}$ =
2.76$^{+1.33}_{-0.81}$ keV and $\Gamma_{2}$ = 1.34$^{+0.36}_{-0.17}$ above
$E_{\rm\,break}$ ($\chi^{2}$/d.o.f = 116/104). The F-test shows an improvement
of the fit with respect to the simple power law ($\chi^{2}$/d.o.f = 126/106)
with a probability of 97.9$\%$, indicating that the broken power-law is the best-fitting model.

\subsection{{\em Swift}-UVOT}

UVOT data in the $v$, $b$, $u$, $w1$, $m2$, and $w2$ filters were analysed
with the \texttt{uvotsource} task included in the \texttt{HEASoft} package
(v6.15) and the 20130118 CALDB-UVOTA release. Source counts were extracted
from a circular region of 5 arcsec radius centred on the source, while
background counts were derived from a circular region with 10 arcsec radius in
a nearby, free region. The observed magnitudes are reported in Table~\ref{UVOT}. Upper limits at 90 per cent confidence level are calculated using the UVOT photometric system when the analysis provided a significance of detection $<$ 3 $\sigma$. 

\begin{table*}
\caption{Observed magnitudes obtained by {\em Swift}-UVOT for PKS 2149$-$306. Upper limits
are calculated when the analysis provided a significance of detection $<$3$\sigma$.}
\begin{center}
\begin{tabular}{cccccccc}
\hline
\multicolumn{1}{c}{Date (UT)} &
\multicolumn{1}{c}{Date (MJD)} &
\multicolumn{1}{c}{$v$} &
\multicolumn{1}{c}{$b$} &
\multicolumn{1}{c}{$u$} &
\multicolumn{1}{c}{$w1$} &
\multicolumn{1}{c}{$m2$} &
\multicolumn{1}{c}{$w2$} \\
\hline
2005-Dec-10 & 53714 & 17.46 $\pm$ 0.14 & 17.76 $\pm$ 0.12 & 17.15 $\pm$ 0.11 & 17.92 $\pm$ 0.14 & 20.22 $\pm$ 0.41 & 19.79 $\pm$ 0.25 \\
2005-Dec-13 & 53717 & 17.56 $\pm$ 0.21 & 17.55 $\pm$ 0.15 & 17.02 $\pm$ 0.12 & 18.12 $\pm$ 0.19 & 19.63 $\pm$ 0.20 & 20.16 $\pm$ 0.44 \\
2009-Apr-23 & 54944 & 17.94 $\pm$ 0.23 & 17.88 $\pm$ 0.13 & 17.30 $\pm$ 0.13 & 18.59 $\pm$ 0.22 & $>$ 19.82 & $>$ 20.37 \\
2009-Apr-29 & 54950 & 17.85 $\pm$ 0.23 & 18.22 $\pm$ 0.16 & 17.85 $\pm$ 0.15 & 18.73 $\pm$ 0.27 & $>$ 19.45 & $>$ 20.21 \\
2009-May-05 & 54956 & 17.32 $\pm$ 0.16 & 17.82 $\pm$ 0.13 & 17.35 $\pm$ 0.14 & 18.52 $\pm$ 0.24 & $>$ 19.69 & $>$ 20.25 \\
2009-May-14 & 54965 & 17.65 $\pm$ 0.17 & 17.99 $\pm$ 0.12 & 17.17 $\pm$ 0.12 & 18.32 $\pm$ 0.19 & $>$ 19.84 & $>$ 20.44 \\
2009-May-23 & 54974 & 17.79 $\pm$ 0.16 & 18.08 $\pm$ 0.11 & 17.34 $\pm$ 0.11 & 18.24 $\pm$ 0.16 & $>$ 20.04 & 20.20$\pm$0.30 \\
2009-May-29 & 54980 & 17.82 $\pm$ 0.23 & 17.71 $\pm$ 0.11 & 17.25 $\pm$ 0.12 & 18.42 $\pm$ 0.19 & $>$ 19.54 & 20.36$\pm$0.39 \\
2010-May-11 & 55327 & 17.84 $\pm$ 0.09 & 18.25 $\pm$ 0.08 & 17.64 $\pm$ 0.08 & 18.38 $\pm$ 0.10 & -- & -- \\
2011-May-07 & 55688 & 17.56 $\pm$ 0.07 & 17.95 $\pm$ 0.10 & 18.12 $\pm$ 0.08 & -- & -- & 20.12 $\pm$ 0.22 \\
2011-Nov-10 & 55875 & 17.92 $\pm$ 0.09 & 18.16 $\pm$ 0.07 & 17.63 $\pm$ 0.08 & -- & -- & -- \\
2013-Dec-16 & 56642 & 17.87 $\pm$ 0.16 & 17.92 $\pm$ 0.10 & 17.22 $\pm$ 0.10 & 18.31 $\pm$ 0.14 & 19.92 $\pm$ 0.30 & 20.24 $\pm$ 0.28 \\
2013-Dec-17 & 56643 & 17.57 $\pm$ 0.18 & 18.14 $\pm$ 0.15 & 17.23 $\pm$ 0.12 & 18.22 $\pm$ 0.18 & $>$ 19.54 & $>$ 20.20 \\
2014-Mar-28 & 56744 & 17.80 $\pm$ 0.20 & 17.87 $\pm$ 0.11 & 17.24 $\pm$ 0.11 & 18.31 $\pm$ 0.17 & $>$ 19.75 & 20.35 $\pm$ 0.36 \\
2014-Apr-18 & 56765 & 17.79 $\pm$ 0.17 & 17.92 $\pm$ 0.10 & 17.24 $\pm$ 0.10 & 18.61 $\pm$ 0.17 & $>$ 20.06 & 20.34 $\pm$ 0.32 \\
\hline
\end{tabular}
\end{center}
\label{UVOT}
\end{table*}

\section{{\em XMM-Newton}: Data Analysis and Results}
\label{XMMData}

{\em XMM-Newton} \citep{jansen01} observed PKS 2149$-$306 on 2001 May 1 for a
total duration of 25 ks (observation ID 0103060401 , PI: Aschenbach). The EPIC
pn was operated in the large-window mode and the EPIC MOS cameras (MOS1 and
MOS2) were operated in the full-frame mode. The data were reduced using the {\em XMM-Newton} Science Analysis System ({\small SAS v14.0.0}), applying
standard event selection and filtering. Inspection of the background light
curves showed that no strong flares were present during the observation, with
good exposure times of 20, 24 and 24 ks for the pn, MOS1 and MOS2,
respectively. For each of the detectors the source spectrum was extracted from
a circular region of radius 30 arcsec centred on the source, and the
background spectrum from a nearby region of radius 30 arcsec on the same chip. All the spectra
were binned to contain at least 20 counts per bin to allow for $\chi^2$
spectral fitting. 

\begin{figure}
\centering
\includegraphics[width=5.2cm, angle=-90]{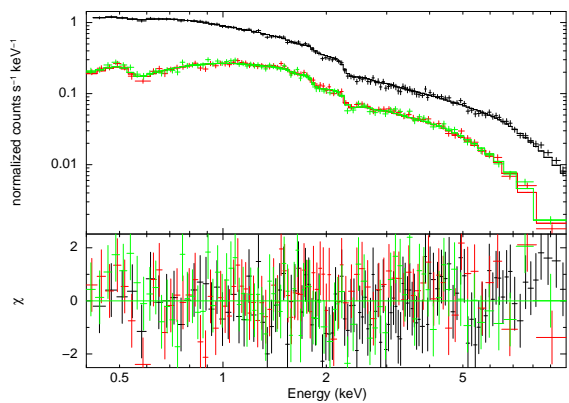}
\caption{EPIC spectra and residuals of PKS 2149$-$306 fitted with a power law model.}
\label{XMM}
\end{figure}

All spectral fits were performed over the 0.3--10~keV energy range using
{\small XSPEC v.12.8.2}. The energies of spectral features are quoted in
the source rest frame, while plots are in the observer frame. All errors are given at the 90\% confidence level.
The data from the three EPIC cameras were
initially fitted separately, but since good agreement was found  ($<5\%$) we
proceeded to fit them together. Galactic absorption was included in all fits using the \texttt{tbabs} model. The
results of the fits are presented in Table~\ref{xmmfits}. As reported also
in \citet{ferrero03} and \citet{bianchin09}, a simple power-law model is
sufficient to describe the data, although some residuals are present
(Fig.~\ref{XMM}). The flux observed by {\em XMM-Newton} in the 0.3--10 keV energy
range is a factor of 2--3 lower than those observed by {\em Swift}-XRT during
2005--2014. 

\noindent A broken power-law does not improve the fit and the associated uncertainties on photon index and flux are larger than those from a fit with a simple power-law (Table~\ref{xmmfits}). In
order to check for the presence of intrinsic absorption, a neutral absorber
at the redshift of the source was added to this model, but it did not improve
the fit quality and thus is not required. Moreover, no Iron line was detected in the spectrum, in agreement with \citet{page04}. The 90\% upper limit on the equivalent width (EW) of a narrow emission line at 6.4~keV is EW$<17$~eV.

\begin{table}
\caption{\small{Summary of fits to the 0.3--10~keV {\em XMM-Newton} spectrum of PKS 2149$-$306. Fits also included absorption fixed at the Galactic value. Flux and E$_{\rm\,break}$ are given in units of erg cm$^{-2}$ s$^{-1}$ and keV, respectively.}}\label{xmmfits}
\begin{center}
\begin{tabular}{lll}
\hline \\ [-4pt]
Model & Parameter & Value \\[4pt]  
\hline\\[-6pt]
Power law & $\Gamma$ & $1.45\pm 0.01$  \\
      & Flux (0.3--10 keV)  &  $(7.9\pm 0.1) \times 10^{-12}$     \\
      &  $\chi^2/\rm{d.o.f.}$          &    375/428  \\
\hline 
Broken Power-law & $\Gamma_1$ & $1.47\pm 0.02$  \\
      & $E_{\rm\,break}$ & $2.4^{+1.3}_{-1.1}$  \\
      & $\Gamma_2$ & $1.42^{+0.03}_{-0.07}$  \\
      & Flux (0.3--10 keV) &  $(7.9\pm 0.2) \times 10^{-12}$     \\
      &  $\chi^2/\rm{d.o.f.}$          &   369/426  \\
\hline
\end{tabular}
\end{center}
\end{table}

\section{{\em NuSTAR}: Data Analysis and Results}
\label{NustarData}

{\em NuSTAR} \citep{harrison13} observed PKS 2149$-$306 with its two coaligned
X-ray telescopes with corresponding focal planes, focal plane module A (FPMA)
and B (FPMB), on 2013 December 17 and on 2014 April 18 for a net exposure time
of 38.5 ks and 44.1 ks, respectively. The level 1 data products were processed
with the {\em NuSTAR} Data Analysis Software (\texttt{nustardas}) package (v1.4.1). Cleaned event files (level 2 data products) were produced and calibrated using standard filtering criteria with the \texttt{NUPIPELINE} task and version
20140414 of the calibration files available in the {\em NuSTAR} CALDB. Spectra of the sources were extracted from the cleaned event files using a
circle of 20 pixel (49 arcsec) radius, while the background was
extracted from two distinct nearby circular regions of 50 pixel
radius. The  ancillary  response  files  were  generated  with  the \texttt{numkarf} task,  applying corrections for the point spread function losses, exposure maps and vignetting. The spectra were rebinned with a
minimum of 20 counts per energy bin to allow for $\chi^{2}$ spectrum fitting. All errors are given at the 90\% confidence level.

\begin{table*}
\caption{Summary of the results for the fits of the 3.0--76~keV {\em NuSTAR} spectra collected on 2013 December 17 and 2014 April 18.}
\begin{center}
\begin{tabular}{cccc}
\hline
\multicolumn{1}{c}{Date} &
\multicolumn{1}{c}{Photon index} &
\multicolumn{1}{c}{Flux 3.0--76 keV} &
\multicolumn{1}{c}{$\chi^2$/d.o.f.}\\
\multicolumn{1}{c}{(UT)} &
\multicolumn{1}{c}{($\Gamma_{\rm\,X}$)} &
\multicolumn{1}{c}{($\times$10$^{-11}$ erg cm$^{-2}$ s$^{-1}$)} &
\multicolumn{1}{c}{}\\
\hline
2013-Dec-17 & $1.37 \pm 0.01$ & $11.5 \pm 0.2$ & 797/805 \\
2014-Apr-18 & $1.46 \pm 0.01$ & $8.2 \pm 0.1$ & 735/747 \\
\hline
\end{tabular}\label{nustar_only}
\end{center}
\end{table*}

By fitting the {\em NuSTAR} spectrum in the 3--76 keV energy range\footnote{We ignored the zero-variance bins in the spectrum, i.e. the 76--79 keV energy range.} a good fit was obtained using a simple power law for both the observations ($\chi^2$/d.o.f. = 797/805 and 735/747), with photon index
$\Gamma_{\rm\,X}$ = 1.37 $\pm$ 0.01 and $\Gamma_{\rm\,X}$ = 1.46 $\pm$ 0.01 (Table~\ref{nustar_only}), that is the same value obtained for $\Gamma_{2}$ with a broken power-law model over the 0.3--76 keV energy range (see Section~\ref{NuSTAR_XRT}). 

\begin{figure}
\centering
\includegraphics[width=5.2cm, angle=-90]{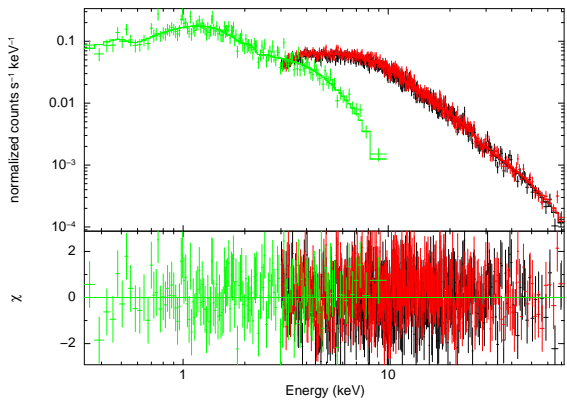}
\caption{{\em NuSTAR} (red and black points) and {\em Swift}-XRT (green points) spectra and residuals of PKS 2149$-$306 collected on 2013 December 16--17, simultaneously fitted with a broken power-law.}
\label{XRTNUSTAR}
\end{figure}

\subsection{Joint {\em NuSTAR} and {\em Swift}-XRT analysis}
\label{NuSTAR_XRT}
 
\begin{table*}
\caption{Summary of the results for the fits of the 0.3--76~keV {\em Swift}-XRT and {\em NuSTAR} spectra collected on 2013 December 16--17 (Obs 1) and 2014 April 18 (Obs 2). All fits also included absorption fixed at the Galactic value.}\label{nustar}
\begin{center}
\begin{tabular}{llll}
\hline\\ [-4pt]
Model& Parameter & Value (Obs 1) & Value (Obs 2) 
 \\[4pt]  
\hline\\[-6pt]

power law & $\Gamma$                       & $1.35\pm 0.01$         & $1.43\pm 0.01$  \\
          & $\chi^2/\rm{d.o.f.}$           & 1013/943               & 983/851\\
\hline 
Broken power-law & $\Gamma_1$              & $0.97^{+0.09}_{-0.07}$   & $0.97^{+0.09}_{-0.07}$ \\
& $\rm{E_{break}}$ (keV)                    & $2.60^{+0.46}_{-0.56}$   & $2.98^{+0.47}_{-0.36}$   \\
& $\Gamma_2$                               & $1.37 \pm 0.01$        & $1.46\pm 0.01$  \\
& $\chi^2/\rm{d.o.f.}$                      &  924/941            &  834/849\\
\hline 
power law  +               & $\Gamma$              		   & $1.36 \pm 0.01$  & $1.45 \pm 0.01$  \\
extra absorber		   & $\rm{N_H^{z}}$ (cm$^{-2}$)              & $1.01^{+0.28}_{-0.24}\times10^{22}$ & $1.36^{+0.39}_{-0.33}\times10^{22}$ \\	
		           & $\chi^2/\rm{d.o.f.}$                  & 966/942 & 895/850\\
\hline

\end{tabular}
\end{center}
\end{table*}

Simultaneously to {\em NuSTAR} observations, {\em Swift}-XRT observations were
performed on 2013 December 16--17 and on 2014 April 18. This allows us to
study the X-ray spectrum of PKS 2149$-$306 over a wide energy range,
i.e. 0.3--76 keV. The results of the simultaneous fits of the {\em NuSTAR} and
{\em Swift}-XRT data are presented in Table \ref{nustar}. The photoelectric
absorption model \texttt{tbabs}, with a neutral hydrogen
column density fixed to its Galactic value (1.63$\times$10$^{20}$ cm$^{-2}$) was included in all
fits. To account for the cross-calibration between {\em NuSTAR}-FPMA, {\em NuSTAR}-FPMB, and {\em Swift}-XRT a constant factor was included in the
model, frozen at 1 for the FPMA spectra and free to vary for the FPMB and XRT
spectra. The X-ray spectrum of the source is not well fitted by a simple power
law model in both the observations ($\chi^2$/d.o.f. = 1013/943 and 983/851,
for the first and second observation, respectively), while a broken power-law
model yielded a good fit  ($\chi^2$/d.o.f. = 924/941 and 834/849). The result
of fitting a broken power-law to the spectrum collected on 2013 December
16--17 is shown in Fig.~\ref{XRTNUSTAR}.  
In this model the power law breaks from a slope of $\Gamma_1 =
0.97^{+0.09}_{-0.07}$ ($\Gamma_1 = 0.97^{+0.09}_{-0.07}$) below $E_{\rm{break}} =
2.60^{+0.46}_{-0.56}~\rm{keV}$ ($2.98^{+0.47}_{-0.36}~\rm{keV}$) to $\Gamma_2
= 1.37 \pm 0.01$ ($\Gamma_{2} = 1.46 \pm 0.01$) for the first (second)
observation (Table~\ref{nustar}). The difference of the cross-calibration for
the FPMB spectra with respect to FPMA spectra is 1--3 per cent, while for the
XRT spectra is less than 10 per cent. These differences become larger (10--30
per cent) when a single power law model is used. By applying an F-test, the
improvement of the fit with a broken power-law is significant with respect to
a single power law, with a probability that the null hypothesis is true of
1.6$\times$10$^{-19}$ and 5$\times$10$^{-31}$ for the first and second
observation, respectively. These results are in agreement with those reported
in \citet{tagliaferri15}. By adding an extra absorption component at the
redshift of the source (\texttt{ztbabs}) to the single power law, the model
provides a good fit to the spectrum, with an equivalent hydrogen column density of $\sim$ 10$^{22}$ cm$^{-2}$, but the quality of the fit is worse than the broken power-law model in both spectra ($\chi^2$/d.o.f. = 966/942 and 895/850; Table~\ref{nustar}). \citet{sambruna07} reported an equivalent hydrogen column density obtained by the fit of {\em Swift} XRT and BAT spectra of 0.25$^{+0.34}_{-0.25}$$\times$10$^{22}$ cm$^{-2}$, that is lower than the values obtained by fitting the {\em Swift}-XRT and {\em NuSTAR} spectra, but their statistics was significantly lower than that presented here.

\section{Discussion}

\subsection{$\gamma$-ray properties}

\begin{figure}
\centering
\includegraphics[width=7.5cm]{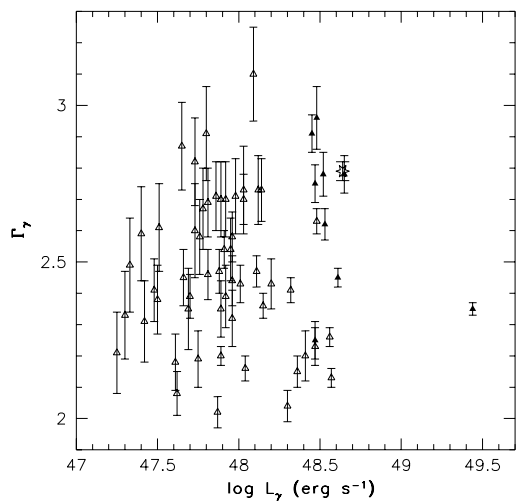}
\caption{$\gamma$-ray photon index vs apparent isotropic luminosity in the 0.1--100 GeV
  energy range for the blazars with $z$ $>$ 2 included in the 3FGL. The
    filled and open triangles represent the objects detected and not detected
    by {\em Swift}-BAT, respectively. The star represents PKS 2149$-$306.}
\label{LATall}
\end{figure}

PKS 2149$-$306 was not associated with a $\gamma$-ray source, either in the LAT
bright source list obtained after three months of {\em Fermi} operation
\citep{abdo09} or in the First
{\em Fermi} LAT source catalogue \citep{abdo10}, indicating that its $\gamma$-ray
activity was low during the first year of {\em Fermi} operation. On the other hand, this FSRQ is
associated with 2FGL J2151.5$-$3021 and 3FGL J2151.8$-$3025 in the Second and
Third {\em Fermi} LAT source catalogues \citep{nolan12,acero15}. The source is not
included in the First {\em Fermi} LAT Catalog of Sources above 10 GeV \citep{ackermann13}.
During the period
2008 August 4--2014 August 4 the $\gamma$-ray spectrum of PKS 2149$-$306 shows significant curvature, well described by a LP model with a spectral slope $\alpha$ = 2.36 $\pm$ 0.05, a curvature parameter around the peak $\beta$ = 0.29 $\pm$ 0.03, and an average flux of (9.7 $\pm$ 0.4)$\times$10$^{-8}$ ph cm$^{-2}$ s$^{-1}$. 

The source showed a significant increase in its $\gamma$-ray flux in 2011
February, and subsequently a strong $\gamma$-ray flare occurred in 2013
January \citep{dammando13}. The flux in 2013 January is about a factor of 8
higher than the average flux estimated over 6 years of {\em Fermi} observations, with a significant change
of the spectral slope ($\alpha$ = 1.99 $\pm$ 0.11) but a similar curvature
parameter ($\beta$ = 0.28 $\pm$ 0.06). This suggests a shift of the IC peak to
higher energies during this flaring activity. In contrast, no significant spectral changes were observed during the flaring activity in 2011
February, when the flux was about a factor of 5 higher than the average
flux. In both flaring episodes the $\gamma$-ray spectrum is well described by
a LP model. Considering the extragalactic background light (EBL) model
discussed in \citet{finke10}, at the redshift of PKS 2149$-$306 the optical
depth should be $\tau$ $\sim 1$ for 50 GeV photons. The maximum photon energy observed from the source during the 2013 flare is 4.8 GeV and is consistent with the current EBL models.

Thirteen FSRQ with $z$ $>$ 2 have been detected by {\em Fermi}-LAT during a
$\gamma$-ray flare up to now. A significant increase of the flux together with a spectral
evolution in $\gamma$ rays was observed for the high-redshift FSRQ TXS
0536$+$145 \citep{orienti14} and S5 0836$+$710 \citep{akyuz13}. In constrast, no significant spectral hardening was observed during the
$\gamma$-ray flares for the high-redshift gravitationally lensed blazar PKS 1830$-$211 \citep{abdo15}.  

During the 2013 flaring activity of PKS 2149$-$306, significant flux variation
by a factor of 2 or more is clearly visible on 12-h and 6-h time-scales, with
the peak of the flare resolved with 6-h binning. In particular, between the
last 6-h bin of January 19 (MJD 56311.875) and the second 6-h bin of January
20 (MJD 56312.625), the flux increases from F$_{1}$ = (88 $\pm$ 40)$\times$10$^{-8}$ to F$_{2}$ = (385 $\pm$ 64)$\times$10$^{-8}$ ph cm$^{-2}$ s$^{-1}$ within $\Delta$t = 12 h, giving a flux doubling time-scale of $\tau_d$ = $\Delta$t $\times$ ln 2/ ln(F$_{2}$/F$_{1}$) $\simeq$ 5.6 h, and an exponential growth time-scale of $\tau_d$/ln2 $\simeq$ 8 h.

\noindent The event horizon light crossing time of a SMBH is t$_{\rm\,lc}$ $\sim$ r$_{\rm\,g}$/$c$ =
G\,M$_{9}$/$c^3$ $\sim 1.4$ $\times$ M$_{9}$ h, where r$_{\rm\,g}$ is the
gravitational radius, M$_{9}$ = (M/10$^{9}$) M$_{\odot}$ is the black hole
(BH) mass, and $c$ the speed of light \citep[e.g.,][]{begelman08}. In the case of
PKS 2149$-$306, with a BH mass of 3.5$\times$10$^{9}$ M$_{\odot}$
\citep{tagliaferri15}, we obtain a $t_{\rm\,lc}$ of $\sim 5 h$, compatible
with the minimum variability detected in the LAT light curve during 2013
January. This short time variability observed in $\gamma$ rays constrains the
size of the emitting region to $R < c t_{var} \delta / (1+z)$ = 2.7
$\times$10$^{15}$ cm \citep[assuming $\delta$ = 14,][]{tagliaferri15}. This
small size of the emitting region should correspond to a small distance from
the central BH, putting the emitting region inside the broad-line region
(BLR). This extremely small size is rather difficult to accommodate in the
`far dissipation' scenario \citep[e.g.,][]{tavecchio10}, where the external
Compton scattering off the infrared photons from the torus is the main
component that produces the high-energy emission, at least during flaring
activity. This is not in contrast to the SED modelling of PKS 2149$-$306
presented in \citet{tagliaferri15}, where low $\gamma$-ray activity
contemporaneous to the {\em NuSTAR} observations was considered. In fact,
different activity states of the same source may have different $\gamma$-ray
emitting region locations. In the case of the 2011 February flare the
statistics are not good enough to determine the flare shape. 

High-redshift blazars tend to be the most luminous AGN due to their
preferential selection by the LAT caused by Malmquist bias \citep{ackermann15}. The daily peak flux
observed on 2013 January 20 is (301 $\pm$ 36)$\times$10$^{-8}$ ph cm$^{-2}$
s$^{-1}$, corresponding to an apparent isotropic $\gamma$-ray luminosity of
(1.5 $\pm$ 0.2)$\times$10$^{50}$ erg s$^{-1}$. On a 6-h time-scale, the flux reached a
peak of (385 $\pm$ 70)$\times$10$^{-8}$ ph cm$^{-2}$ s$^{-1}$, corresponding
to an apparent isotropic $\gamma$-ray luminosity of (1.9 $\pm$ 0.3)$\times$10$^{50}$erg
s$^{-1}$. As a comparison, the average $\gamma$-ray luminosity over 6 years of {\em Fermi} operation is
4.4$\times$10$^{48}$ erg s$^{-1}$. The peak values are comparable to the highest luminosity observed from FSRQ so
far \citep[i.e., 3C 454.3 and PKS 1830$-$211;][]{ackermann10, abdo15} and a factor
of two higher than the peak luminosity observed from TXS 0536$+$135, that is the most distant $\gamma$-ray flaring blazar observed by {\em Fermi}-LAT up to now \citep{orienti14}. In Fig.~\ref{LATall}, we compare this value
with the $\gamma$-ray luminosity of all blazars with $z$ $> 2$ included in the
3FGL. We consider the high-redshift blazars detected also by {\em Swift}-BAT in hard X-rays. In particular, in the 70-month {\em Swift}-BAT catalogue \citep{baumgartner13} there are 17 blazars with redshift $z$ $>$ 2. The filled and open triangles in Fig.~\ref{LATall} represent the blazars detected and not detected by {\em Swift}-BAT, respectively. The $\gamma$-ray luminosity, $L_{\gamma}$, is computed following \citet{ghisellini09}:

\begin{equation} 
L_{\gamma} = 4 \pi d_{\rm L}^2 \frac{S_{\gamma}}{(1+z)^{2-\Gamma_{\gamma}}}
\label{eq_lum}
\end{equation}

\noindent where $S_{\gamma}$ is the energy flux between 100 MeV and 100 GeV, and $\Gamma_{\gamma}$ is the
photon index.
 
All these high-redshift blazars are FSRQ, with the exception of SDSS J145059.99+520111.7, PMN
J0124-0624, MG4 J000800+4712, and PKS 0437-454 classified as BL Lac
objects.
Considering the average luminosity, PKS 2149$-$306 is the third brightest
object among the high-redshift blazars
detected by LAT, after PKS 1830$-$211 and PKS 0537$-$286. 

\begin{figure}
\centering
\includegraphics[width=7.5cm]{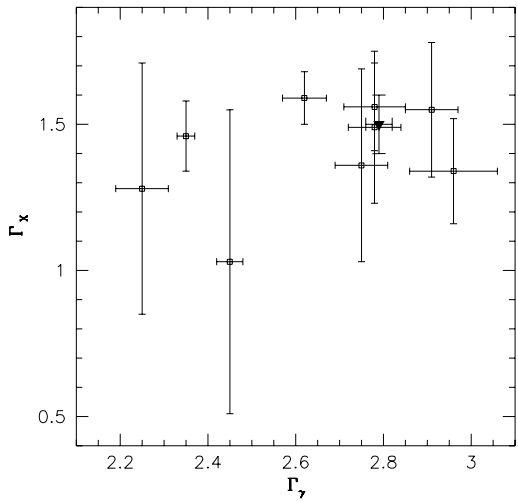}
\caption{X-ray photon index from {\em Swift}-BAT vs $\gamma$-ray photon index from {\em Fermi}-LAT of the 10 high-redshift blazars detected by both
 instruments. The filled upside down triangle represents PKS 2149$-$306.}
\label{BAT_LAT}
\end{figure}

By considering the blazars in the {\em Swift}-BAT catalogue, we note that all the high-redshift blazars detected by both {\em Fermi}-LAT and {\em Swift}-BAT have $L_{\gamma}$ $ >$ 2$\times$10$^{48}$ erg s$^{-1}$, suggesting that only the most luminous $\gamma$-ray blazars are detected by both instruments. Most of the LAT sources detected by BAT, including PKS 2149$-$306, have a soft $\gamma$-ray photon index $\Gamma_{\gamma}$ $> 2.5$. This corresponds in hard X-rays to a photon index $\Gamma_{\rm\,X}$ $< 1.6$ (see Fig.~\ref{BAT_LAT}). This result confirms that the detection of these high-redshift blazars strongly depends on the position of their IC peaks. According to the blazar sequence \citep{fossati98}, as the bolometric luminosity of a blazar increases the synchrotron and IC peak moves to lower frequencies. Considering that high-redshift FSRQ are powerful blazars with high bolometric luminosity, their IC peak is usually expected in the 1--10 MeV energy range, below the energy range covered by {\em Fermi}-LAT \citep[see e.g.,][]{tagliaferri15}. However, during strong flaring activity the IC peak may shift to higher energies as in the case of
the 2013 flare from PKS 2149$-$306. 

Twenty-eight high-redshift blazars detected by {\em Fermi}-LAT have $L_{\gamma}$ $>$ 10$^{48}$ erg s$^{-1}$. The vast majority of them have a BH mass $>$ 10$^{9}$ M$_{\odot}$ \citep{ghisellini09,ghisellini10,ghisellini11,ghisellini14b}, confirming that the most powerful blazars have the heaviest BH \citep{ghisellini13b}. In particular, PKS 2149$-$306 has a BH mass of 3.5$\times$10$^{9}$, as estimated by \citet{tagliaferri15}.

\subsection{X-ray properties}

\begin{figure}
\centering
\includegraphics[width=7.5cm]{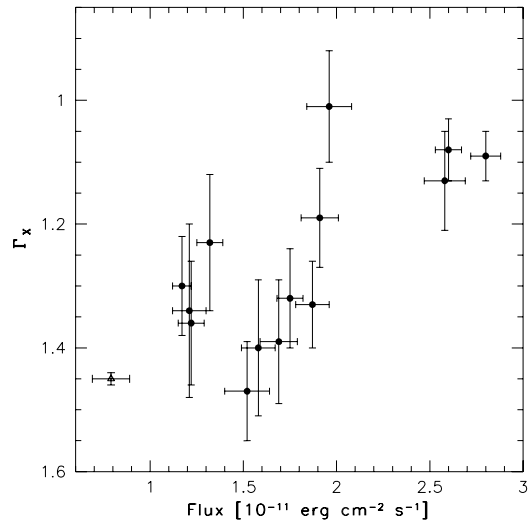}
\caption{X-ray photon index vs 0.3--10 keV flux of PKS 2149$-$306 during 2001--2014. Filled circles are {\em Swift}-XRT observations; the open triangle represents the {\em XMM-Newton} observation.}
\label{XRay_photon}
\end{figure}

We investigated the X-ray properties of PKS 2149$-$306 by means of {\em
  Swift}-XRT, {\em XMM-Newton}, and {\em NuSTAR} observations. The X-ray
spectrum collected by {\em XMM-Newton} in 2001 is quite well modelled by a
simple power law with a photon index of $\Gamma_{\rm\,X}$ = 1.45 $\pm$ 0.01
and a 0.3--10 keV flux 0f 7.9$\times10^{-12}$ erg cm$^{-2}$ s$^{-1}$. During 2005
December--2014 April, {\em Swift}-XRT observed the source with a 0.3--10 keV flux in the range (1.2--2.8)$\times$10$^{-11}$ erg cm$^{-2}$ s$^{-1}$, with a photon index varying between 1.0 and 1.5. 
Fig.~\ref{XRay_photon} shows the X-ray photon index estimated from {\em Swift}-XRT and {\em XMM-Newton} observations as a function of the X-ray flux in the 0.3--10 keV range: despite the large errors, a hint of hardening of the spectrum with the increase of the flux is observed.

\begin{figure}
\centering
\includegraphics[width=7.5cm]{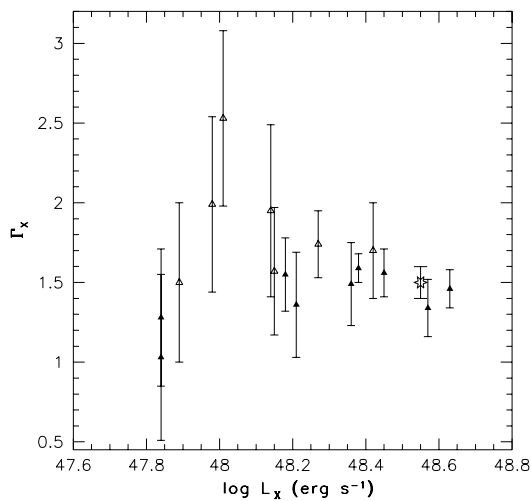}
\caption{X-ray photon index vs X-ray apparent isotropic luminosity in the 14--195 keV
  energy range for the blazars with $z$ $>$ 2 detected by {\em Swift}-BAT. The filled and open triangles represent the objects detected and not detected by {\em Fermi}-LAT, respectively. The star represents PKS 2149$-$306.}
\label{BAT_all}
\end{figure}

Unfortunately the {\em Swift} observations did not cover the $\gamma$-ray flaring periods detected in 2011 February and 2013 January, preventing us from investigating the X-ray behaviour during these $\gamma$-ray flaring events.

The {\em {NuSTAR}} spectra collected in the 3--76 keV energy range during 2013 December 17 and 2014 April 18 are well fitted by a simple power law
  with photon index $\Gamma_{\rm\,X}$ = 1.37 $\pm$ 0.01 and $\Gamma_{\rm\,X}$ = 1.46 $\pm$ 0.01, respectively. The two simultaneous observations of PKS 2149$-$306 by {\em Swift}-XRT and {\em NuSTAR} showed that the broad band X-ray spectrum is well described by a broken power-law model, with a very hard spectrum ($\Gamma_{1}$ $\sim$1) below the break energy, at E$_{\rm\,break}$ = 2.5--3.0 keV, and $\Gamma_{2}$ = 1.37 $\pm$ 0.01 and 1.46 $\pm$ 0.01 above the break energy. {\em Swift}-BAT and {\em BeppoSAX} observed a photon index $\Gamma_{\rm\,X}$ = 1.50 $\pm$ 0.10 \citep{baumgartner13} and
$\Gamma_{\rm\,X}$ = 1.40 $\pm$ 0.04 \citep{elvis00} in the 14--195 keV and 20--200 keV energy range, respectively, in agreement with the photon index $\Gamma_{2}$ obtained by the two {\em Swift}-XRT and {\em NuSTAR} joint fits. The 3--76 keV flux varied by about 40 per cent between the first and second {\em NuSTAR} observation. At the same time the 0.3--10 keV flux varied by less than 10 per cent. In the same way the photon index below the break energy did not change, while the photon index above the break energy was harder when the source was brighter. 

In several high-redshift ($z$ $> 4$) blazars a steepening of the soft X-ray spectrum has been observed \citep[e.g.,][and references therein]{yuan06}. This steepening may be due to either an excess of absorption in the soft X-ray part of the spectrum or due to an intrinsic curvature of the electron energy distribution responsible for the X-ray emission. In PKS 2149$-$306 this feature was observed below 3 keV, and an improvement of the fit is observed when an extra absorber at the redshift of the source (N$_{\rm\,H}^{z}$ $\sim$ 10$^{22}$ cm$^{-2}$) is added to the simple power law model. However, this improvement is not as good as when we use a broken power-law model. The AGN may be surrounded by a dense
plasma in form of a wind or an outflow \citep[e.g.,][]{fabian99}. This intervening material may be responsible for the extra absorption observed in high-redshift quasars \citep[e.g.,][]{vignali05}. 
However, in contrast to radio-loud quasars, it is  unlikely that for a
blazar like PKS 2149$-$306 such a large gas column density in the line of
sight is not removed by the relativistic jet that should be well aligned with the line of sight. Moreover, with a hydrogen column density of about
10$^{22}$ cm$^{-2}$ obtained by fitting the X-ray spectrum the corresponding
optical obscuration \citep[see e.g.,][]{guver09} would be very high ($A_{V}$
$\sim 10$) and the source would not be detectable in optical-UV with the
short exposures of the {\em Swift} observations. This problem may be solved by invoking a high ionization state of the gas, but it is not possible to have conclusive evidence from its X-ray spectrum. In fact, due to the redshift of this source the most important spectral features used as a diagnostic are out of the energy range covered by {\em Swift} and {\em XMM-Newton}.

The most likely explanation of the steepening of the spectrum is that the soft
X-ray emission is produced by external Compton radiation from the electrons at the lower
end of the energy distribution \citep[e.g.,][]{tavecchio07}. This is in
  agreement with the stable photon index below 3 keV and the change of the
  photon index above $\sim 3$ keV, where the emission is produced by the most
  energetic electrons. In this context the lack of a clear steepening in the
  X-ray spectrum collected by {\em XMM-Newton} in 2001 may be related to the
  lower flux with respect to that estimated during the {\em Swift}-XRT and
  {\em NuSTAR} observations. The intrinsic curvature in soft X-rays would be
  less evident during the low activity of the source. The situation may be
  more complex due to the possible presence in the X-ray band also of the
  synchrotron self-Compton emission (SSC). However, the relative importance of
  the SSC component should decrease with the luminosity of the source
  \citep[e.g.,][]{ghisellini98}, and therefore should be negligible in
  powerful FSRQ such as PKS 2149$-$306. Moreover, \citet{celotti07} proposed the
  presence of a spectral component in X-rays produced by the Comptonization of
  ambient photons by cold electrons in the jet approaching the BLR. However,
  the direct detection of such a component has remained elusive. It is worth
  nothing that the broad-band spectrum of PKS 2149$-$306 including the {\em Swift}-XRT and {\em NuSTAR} data are better fitted by considering the
  emitting region outside the BLR \citep[e.g.,][]{tagliaferri15}, where the contribution of the bulk Comptonization should be negligible.

Considering the blazars included in the 70-month {\em Swift} BAT catalogue with a redshift $z$ $> 2$, 7 out of 17 have not been detected by
{\em Fermi}-LAT so far. Only sources with a $\Gamma_{\rm\,X} < 1.6$ have been detected in $\gamma$ rays, while no dependence on the X-ray luminosity seems to be evident (Fig.~\ref{BAT_all}). This is confirmed by the fact that the average photon index of the FSRQ detected by LAT,
$<\Gamma^{\rm\,LAT}_{\rm\,X}>$ = 1.42 $\pm$ 0.09 is quite different from that
of the sources detected by BAT and not by LAT,
$<\Gamma^{\rm\,no\_LAT}_{\rm\,X}>$ = 1.85 $\pm$ 0.17. Seven blazars with $z$
$> 3$ have been detected by {\em Swift}-BAT, and another one, IGR
J12319$-$0749, by {\em INTEGRAL}-IBIS \citep{bassani12}. In addition, two blazars at redshift $z > 5$ have been detected by {\em NuSTAR}. Only two of these ten blazars have been detected by {\em Fermi}-LAT: PKS 0537$-$286 \citep[e.g.,][]{bottacini10} and TXS 0800$+$618 \citep[e.g.,][]{ghisellini10}, confirming that the $\gamma$-ray energy range is not ideal for detecting blazars at redshift $> 3$. Among the FSRQ detected by both BAT and LAT, PKS 2149$-$306 is the third most luminous after PKS 1830$-$211 \citep[e.g.,][]{abdo15} and B2 0743$+$25 \citep[e.g.,][]{ghisellini10}. 

\subsection{Optical and UV properties}

In powerful high-redshift blazars, such as PKS 2149$-$306, the synchrotron peak is shifted to the mm-regime, leaving the
thermal emission from the accretion disc as the dominant contribution in the
optical-UV part of the spectrum \citep[e.g.,][]{tagliaferri15}. This accretion disc emission is not expected
to vary on short time-scales. 
During 2005--2014 the difference between the maximum and minimum magnitude
observed by {\em Swift}-UVOT is 0.6, 0.7, 1.1, 0.8, 0.6, and 0.6 mag (corresponding to a variation of the flux density of 1.5,
1.8, 2.8, 2, 1.5, 1.5) from the $v$ to the $w2$ band. No significant variability
was observed on a time-scale of a few days, in agreement with thermal emission from an
accretion disc.

\section{Summary}

In this paper we discussed the $\gamma$-ray and X-ray properties of the high-redshift FSRQ PKS 2149$-$306 by means of {\em Fermi}-LAT, {\em NuSTAR}, {\em XMM-Newton}, and {\em Swift} data. We summarize our main conclusions as follows:

\begin{itemize}

\item PKS 2149$-$306 showed a significant increase in its $\gamma$-ray activity in 2011 February and 2013 January. During the 2013 flare the flux
  increase was accompanied by a significant change of the spectral slope, not observed during the 2011 flare.

\item During the 2013 $\gamma$-ray flaring activity significant flux
  variations are observed on a 6-hr time-scale, compatible with the light
  crossing time of the event horizon of the super-massive black hole (SMBH). On 2013 January 20, the source reached a daily $\gamma$-ray peak
  flux of (301 $\pm$ 36)$\times$10$^{-8}$ ph cm$^{-2}$ s$^{-1}$, up to (385
  $\pm$ 70)$\times$10$^{-8}$ ph cm$^{-2}$ s$^{-1}$ on a 6-hr time-scale. These values correspond to an apparent isotropic
  $\gamma$-ray luminosity of (1.5 $\pm$ 0.2)$\times$10$^{50}$ and (1.9 $\pm$ 0.3)$\times$10$^{50}$ erg
  s$^{-1}$, respectively, comparable to the highest values observed from FSRQ up to now.

\item The average $\gamma$-ray luminosity of PKS 2149$-$306 over 6 years of {\em Fermi} operation is
4.4$\times$10$^{48}$ erg s$^{-1}$. This is the third brightest blazar with
$z$ $> 2$ detected by LAT, after PKS 1830$-$211 and PKS 0537$-$286.

\item All high-redshift blazars detected by both {\em Fermi}-LAT and {\em
    Swift}-BAT have a $L_{\gamma}$ $ > 2$$\times$10$^{48}$ erg s$^{-1}$,
  suggesting that only the most luminous $\gamma$-ray blazars are detected by
  both instruments. Like most of the LAT blazars detected by
  BAT, PKS 2149$-$306 has a soft $\gamma$-ray photon index $\Gamma_{\gamma}$ $> 2.5$. This
  corresponds to a photon index $\Gamma_{\rm\,X}$ $< 1.6$ in hard X-rays.

\item Among the FSRQ with $z$ $> 2$ detected by both BAT and LAT, PKS 2149$-$306 is
  the third most luminous in hard X-rays after PKS 1830$-$211 and B2 0743$+$25.

\item The broad band X-ray spectrum of PKS 2149$-$306 observed by {\em Swift}-XRT and {\em NuSTAR} is well described by a
broken power-law model, with a very hard spectrum ($\Gamma_{1}$ $\sim 1$)
below the break energy, at E$_{\rm\,break}$ = 2.5--3.0 keV, and $\Gamma_{2}$
$\sim 1.4$--1.5 above the break energy. 

\item The steepening of the spectrum below $\sim 3$ keV could be due to
  the fact that the soft X-ray emission is produced by the low-energy tail of
  the relativistic electrons producing IC emission. This is in agreement with
  the small variability amplitude and the lack of spectral changes observed in
  that part of the X-ray spectrum between the two {\em NuSTAR} and {\em Swift}
  joint observations. An extra absorption due to material surrounding the SMBH
  is unlikely because the relativistic jet should efficiently remove the gas
  along the line of sight. Moreover, this extra absorption should correspond to a very large extinction in optical and UV, in contrast to the detection of the source by {\em Swift}-UVOT.

\item {\em Fermi}-LAT and {\em Swift}-BAT observations are confirming that the hard X-ray band is more effective in selecting bright FSRQ at $z$ $>$ 3 \citep[see e.g.,][]{ghisellini10, ajello12}, while the $\gamma$-ray band is very effective up to $z$ = 2 \citep[see e.g.,][]{ackermann15}.

\end{itemize}

Further multiwavelength observations of PKS 2149$-$306 will be important for shedding light on the properties of high$-$z blazars. In
particular, simultaneous optical-to-X-ray observations during a $\gamma$-ray flaring activity will allow us to compare its broad-band spectral energy distribution during both low and high activity states, constraining the emission mechanisms at work.
                 
\section*{Acknowledgements}

The {\em Fermi} LAT Collaboration acknowledges generous ongoing
support from a number of agencies and institutes that have supported
both the development and the operation of the LAT as well as
scientific data analysis.  These include the National Aeronautics and
Space Administration and the Department of Energy in the United
States, the Commissariat \`a l'Energie Atomique and the Centre
National de la Recherche Scientifique / Institut National de Physique
Nucl\'eaire et de Physique des Particules in France, the Agenzia
Spaziale Italiana and the Istituto Nazionale di Fisica Nucleare in
Italy, the Ministry of Education, Culture, Sports, Science and
Technology (MEXT), High Energy Accelerator Research Organization (KEK)
and Japan Aerospace Exploration Agency (JAXA) in Japan, and the
K.~A.~Wallenberg Foundation, the Swedish Research Council and the
Swedish National Space Board in Sweden. Additional support for science
analysis during the operations phase is gratefully acknowledged from
the Istituto Nazionale di Astrofisica in Italy and the Centre National
d'\'Etudes Spatiales in France.

Part of this work was done with the contribution of the Italian Ministry of
Foreign Affairs and Research for the collaboration project between Italy and
Japan. We thank the {\em Swift} team for making these observations possible, the
duty scientists, and science planners. This  research  has made use of the
{\em NuSTAR} Data Analysis Software (NuSTARDAS) jointly  developed  by  the  ASI
Science  Data Center (ASDC,  Italy)  and  the  California  Institute  of
Technology (USA). We thank Eugenio Bottacini, Luca Baldini, and Jeremy Perkins for useful comments and suggestions.

\end{document}